# La variabilité des inondations prise la main dans le sac !

## Drawing a better understanding of flood quantiles from a bag

POULARD Christine[1], RENARD Benjamin[2], GONZALEZ-SOSA Enrique[3], CAILLOUET Laurie[4]

[1]INRAE, UR RiverLy, Centre de Lyon-Grenoble et [2] Centre de Aix-en-Provence (prénom.nom@inrae.fr), [3]Université de Queretaro, Mexique (egs@uaq.mx)
[4]Association « EauDyssée » (eaudyssee.asso@gmail.com)

## RÉSUMÉ

La «crue centennale » s'invite souvent dans les journaux, mais les notions d'aléa d'inondation sont finalement complexes à appréhender. Nous présentons d'abord une animation 'tout public', pour rendre les quantiles de débit ou de pluie plus concrets au moyen d'un sac de billes de plusieurs couleurs, codant des classes de périodes de retour. Discuter de l'analogie entre des tirages dans ce sac et la survenue de crues aidera à expliquer qu'il faut raisonner (i) sur toute la gamme des débits, dont le « centennal », (ii) sur le moyen ou le long terme ( tirages successifs), ce qui change la perception de l'aléa, (iii) en tenant compte de la variabilité. Un code qui automatise les tirages confirme que les probabilités empiriques rejoignent les probabilités théoriques, mais illustre aussi des résultats d'analyse combinatoire pas toujours intuitifs : sur des séries de 100 années, en moyenne un quart subissent au moins deux crues de débit supérieur au débit centennal... Le deuxième code, Sample2Gumbel, est un outil pour l'enseignement, qui approfondit l'expérience en générant des valeurs de débit maxima annuels. Il crée un graphique comportant (i) la « vraie » loi, utilisée pour générer l'échantillon, (ii) cet échantillon tracé en période de retour empirique, notion usuelle mais à ne pas sur-interpréter, (iii) la distribution ajustée sur l'échantillon. Relancer une nouvelle série permet de constater la variabilité due au hasard, et ajouter des tirages à la série en cours montre comment l'incertitude évolue. Cet outil se prête à des améliorations : introduire davantage de lois et ajouter une étape de calcul de dommages.

## ABSTRACT

The « 100-year flood » is commonly used, for instance in newspapers, but flood hazard assessment is more complex than it seems. We first describe an animation entitled « bag of floods » to make flood quantiles more concrete, using marbles whose colour corresponds to a class of return period. Discussing the analogies and differences between drawing a marble from the bag and the next annual flood make it easier to explain that flood hazard assessment  (i) must not be focussed on the « 100-yr flood », (ii) is often expressed as a probability over one given year, but for planning it should be estimated over a much longer duration (like successive draws from the bag) and (iii) variability is significant and matters. Scripts allowing to simulate long series of draws confirm that empirical probabilities get close to theoretical probabilities, but also illustrate less intuitive results : on average one quarter of 100 successive draws, contains two floods or more with a discharge exceeding the « 100-yr discharge». To go further, Sample2Gumbel is a teaching tool drawing annual maxima discharges. It compares on a graph (i) a "real distribution", coded in the script and used to draw a sample, (ii) the sample expressed with respect to "plotting position", expressed as a return period but which is in fact a crude estimation to allow plotting, (iii) the distribution fitted on the sample. This demo tool illustrates the variability of different tries, with samples of the same length, and shows how uncertainty evolves with the sample size. To improve it, more distributions could be included, and damage estimation could be added.

## MOTS CLÉS

Crue centennale ; hydrologie statistique ; quantiles de crue ; vulgarisation

Nous présentons deux petits projets motivés par le constat que la notion de « crue centennale », et plus généralement de crues « de période de retour donnée » sont plus complexes qu'il n'y paraît à appréhender, y compris par des étudiants. Est-ce que l'on peut parler de quantiles « de crue » ou se restreindre à la notion de quantiles de débit ? Comment les estime-t-on ? Dans les deux cas, nous insistons sur la variabilité du phénomène, une dimension qui mérite d'être mieux prise en compte.

# 1 UNE EXPÉRIENCE GRAND PUBLIC : « LES CRUES LA MAIN DANS LE SAC »

Cette démarche a été initiée pour une « Fête de la Science », dans le hall hydraulique d'Inrae à Villeurbanne. En introduction, les intervenants définissaient la « crue centennale » en utilisant une analogie avec un tirage aléatoire dans un sac contenant 100 billes dont une rouge. Inviter les participants à tirer une bille d'un sac capte leur attention et facilite les échanges. Avec des tirages successifs, on raisonne sur la probabilité d'inondation sur plusieurs années. L'animation a été étoffée et décrite sur le wiki d'un gitlab[1], plutôt à l'attention de pairs intéressés par notre animation et/ou nos codes. Enfin, un billet « tout public » a été rédigé conjointement avec l'association EauDyssée pour diffuser plus largement les principaux messages (https://www.eaudyssee.org/la-crue-centennale-prise-la-main-dans-le-sac/ ).

## 1.1 Que représentent les billes et leur code couleur ?

Il faut une métrique pour « classer » les inondations et faire des statistiques ; pour un débordement de cours d'eau ce sera le **débit de la rivière**, et pour du ruissellement ce sera plutôt l'**intensité de la pluie**. On décrit ici la démarche pour les crues, en admettant qu'une crue est « centennale » quand son débit maximal est centennal, ce qui facilite les analyses statistiques mais méritera d'être rediscuté en conclusion car certaines inondations sont plutôt remarquables par leur durée et/ou leur volume.

Commenter les définitions proposées permet de tester l'aisance des participants avec les notions de période de retour et de probabilité. La définition la plus fréquente d'un débit centennal est qu'il « survient une fois tous les 100 ans » ; il est en fait atteint **ou dépassé** en moyenne une année sur 100, soit une « chance » sur 100 d'être dépassé une année donnée... Tirer une bille rouge correspond donc à une crue **au moins centennale**...

La« centennale » mérite-t-elle d'accaparer toute l'attention ? Selon les endroits, les inondations commencent avant ou après le niveau centennal... Les dégâts ne sont pas causés que par la centennale : il est donc plus judicieux de raisonner sur une gamme de débits. Pour apporter cette nuance, notre sac se compose de 90 billes noires (débit inférieur au décennal), 8 billes vertes ( entre le décennal et le cinquantennal), une bille bleue (plus que cinquantennal, moins que centennal), et une rouge (centennale ou plus) : les couleurs codent donc des **classes de périodes de retour**. Selon le public, on annoncera ou l'on construira avec eux la composition du sac.

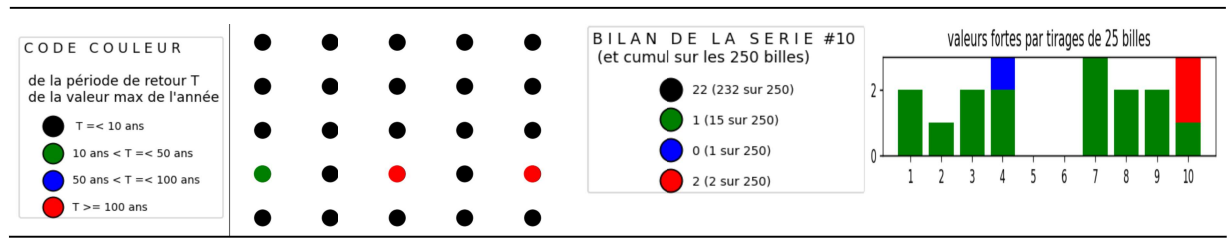

Figure 1: de gauche à droite : captures d'écran de « sacdecrues.py » : code couleur , série de 25 tirages, compteurs de cette série et du cumul des 10 séries tirées, rappel des billes « rares » des 10 tirages

## 1.2 Intérêt de tirages successifs

On ne construit pas une habitation ou un ouvrage hydraulique pour une année, mais pour plusieurs décennies. Mace (2021) propose de raisonner sur la durée d'un crédit immobilier ; à défaut d'être forcément le bon horizon temporel, cette suggestion est pertinente car très concrète pour le grand public. Pour se représenter l'aléa d'inondation sur ces N années, tirons N billes successivement, avec remise.

---

1 URL actuelle (juillet 2023) : https://forgemia.inra.fr/christine.poulard/hydrodemo-01-sacdecrues

Plusieurs codes sont disponibles pour simuler des tirages bille par bille, ou par séries. Une version en javascript est facile à distribuer car elle ne nécessite qu'un navigateur. La version en Python offre plus de possibilités, et affiche des bilans détaillés par série et par cumul de séries (10 séries de 25 tirages sur la figure 1). En enchaînant des tirages, ou en augmentant la taille des séries, les ratios se rapprochent des probabilités théoriques. Mais multiplier les tirages montre aussi leur **variabilité,** illustrée par le graphique du bas de la Figure 1 qui conserve une trace du nombre de « billes rares » pour chaque série. **L**es séquences « atypiques » ne sont pas anormales, elles font partie des réalisations possibles.

Comme application pratique, regardons l'emprise d'une « crue centennale » sur une carte d'aléa inondation : une maison située pile à sa la limite serait inondée en moyenne tous les 100 ans (Poulard, et al. 2023). Quelle est sa probabilité d'être inondé au moins une fois en 100 ans ? Selon le public visé, on abordera ce calcul empiriquement avec « sacdecrues.py », ou analytiquement : la probabilité d'avoir sur N années données exactement k crues qui dépassent le niveau de la crue de période de retour T se calcule par la formule :

$$C_k^N \cdot \left(\frac{1}{T}\right)^k \cdot \left(1-\left(\frac{1}{T}\right)\right)^{N-k} \qquad \text{(Eq. 1)}$$

Un tableur, ou un troisième code fourni[2] donne la réponse : 37 % de « chances » de ne pas être inondé dans une fenêtre de 100 années, 37 % de l'être une fois exactement… et 26 % de l'être deux fois ou plus.

### 1.3 Enseignement de ces expériences

Cette démarche met en évidence la variabilité du phénomène, et invite à réfléchir :

- interpréter une probabilité de crue pour une prise de décision paraît simple ; le raisonnement conduit sur N années montre que l'aléa est en fait difficile à appréhender sur le moyen et long terme.
- le hasard joue grand rôle dans la genèse des inondations… et estimer la relation entre débit et période de retour à partir de quelques années d'observation sera donc délicat ; c'est l'objet du point suivant.

## 2 UN CODE DE DEMONSTRATION POUR L'ENSEIGNEMENT : SAMPLE_2_GUMBEL

Le code Sample_2_Gumbel[3] est dans la continuité du premier, mais à destination d'étudiants du supérieur ou de praticiens. C'est un démonstrateur conçu pour manipuler les notions d'échantillon, de période de retour empirique, d'ajustement statistique avec son incertitude associée, et pour illustrer la variabilité et ses conséquences sur les estimations.

Ce code ne permet actuellement de travailler qu'avec une loi de Gumbel, à 2 paramètres, très utilisée en hydrologie, pour représenter la relation entre les débits Maxima Annuels, notés QXA et la période de retour T (e.g. cours de Renard, en référence). Il pourrait bien entendu être amélioré en ajoutant d'autres lois, et même en incluant un calcul de dommages ; nous sommes ouverts à toutes collaborations.

### 2.1 Tirage d'un échantillon et représentation en période de retour empirique

Une loi de Gumbel est définie dans le code : on suppose qu'elle représente une relation QXA(T) pour un bassin versant donné ; ses deux paramètres sont modifiables dans le code. Elle est tracée en tirets orange sur l'interface ( Figure 2, droite). Le menu propose : le tirage d'un échantillon de 10 valeurs selon la loi définie dans le code, l'ajout de 10 nouvelles valeurs au tirage en cours, la bascule de l'axe des abscisses soit en période de retour soit en probabilité.

Comme on connaît la loi, on peut en déduire aussitôt la période de retour, et représenter l'échantillon selon le même code couleur que celui du« sac de crues » (vignette du haut). Dans une vraie étude on ne connaît pas QXA(T), c'est ce que l'on cherche à estimer à partir des années d'observation disponibles.

Sur le graphique du bas, échantillon trié est représenté en fonction d'une période de retour empirique ou de la probabilité au dépassement. Cette visualisation classique est pratique mais ne doit pas être surinterprétée : quand l'axe des abscisses est exprimé en probabilité, on vérifie que les fréquences empiriques sont simplement obtenues en découpant régulièrement l'intervalle [0,1] . Il existe plusieurs formulations, qui diffèrent par leurs paramètres, notés a et b. Relancer une nouvelle série de 10 tirages confirmera qu'elles ne dépendent pas du tout des valeurs de l'échantillon, simplement du nombre d'années et du rang de chaque observation après tri.

---

2  ProbaCruesMaxAn_SurNannees.py : pour T et N donnés, trace une figure avec plusieurs k

3  URL actuelle (juillet 2023) : https://gitlab.irstea.fr/hydrotools_demosandprocessing/sample2gumbel

## 2.2 Ajustement d'une loi de Gumbel et calcul des incertitudes

L'ajustement consiste ici à estimer les paramètres de la loi de manière à ce que la moyenne et l'écart-type de la loi soient égaux à ceux de l'échantillon ; il est représenté en trait bleu. Un ajustement conduit sur un tirage avec des débits particulièrement forts surestimera les quantiles, et inversement des débits modestes amèneront une sous-estimation. L'intervalle de confiance est calculé (cf cours de Renard) et tracé en jaune. La figure 2 illustre que l'intervalle de confiance s'accroît avec la période de retour ; en augmentant le nombre d'observations on vérifie que l'intervalle de confiance s'affine pour une période de retour donnée, et que l'ajustement aura tendance à se rapprocher de la loi de départ.

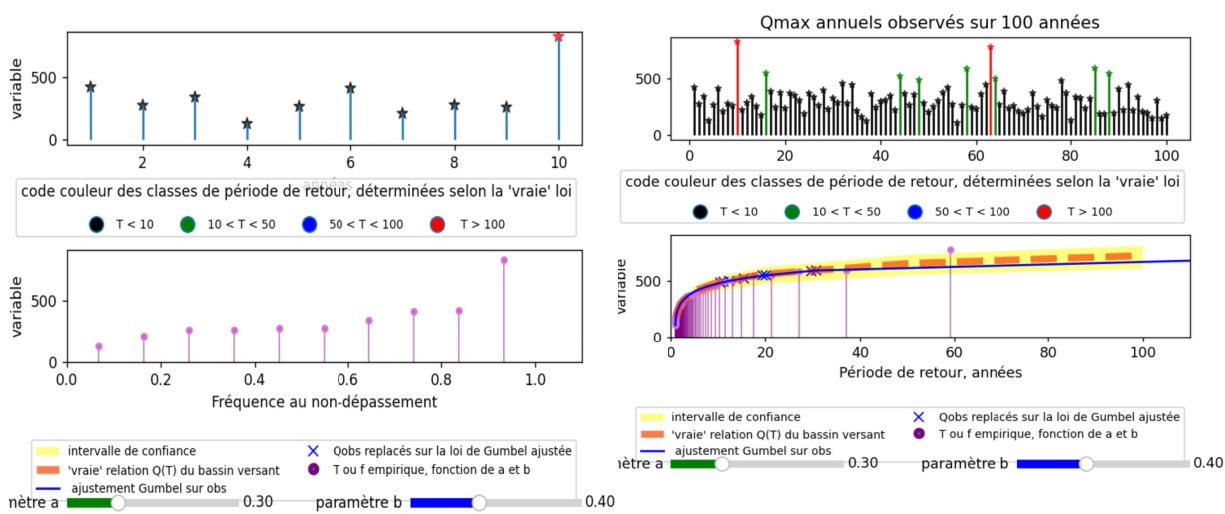

Figure 2 : interface, avec en en haut la représentation chronologique des tirages et en bas représentation en fréquence (gauche) ou en période de retour (droite). A gauche : tirage de 10 valeurs ; à droite échantillon a complété à 100 valeurs ;

## 2.3 Enseignement de ces expériences

L'application permet simplement d'illustrer des notions de cours ; les fonctionnalités ont été implémentées en fonction des difficultés constatées : signification de la période de retour empirique, limite de validité de l'ajustement pour des périodes de retour rares…

Les pistes d'amélioration sont nombreuses : il serait intéressant de coder une loi de comportement à trois paramètres et de chercher à ajuster une loi à deux paramètres, ou d'implémenter une loi exponentielle qui convient à des échantillons de débit sup-seuil. Nous avons aussi testé l'ajout d'une étape d'estimation de dommages à partir d'une relation débit-dommages établie au préalable. Cette fonctionnalité serait à affiner en fonction des objectifs, par exemple illustrer la variabilité des dommages empiriques sur une durée donnée.